\documentclass{article}
\usepackage{spconf, amsmath, graphicx}
\usepackage{amssymb}
\usepackage{multirow}
\usepackage{balance}
\usepackage{xcolor}
\usepackage{url}

\DeclareMathOperator*{\argminX}{arg\,min} 

\title{Autonomous Polycrystalline Material Decomposition for Hyperspectral Neutron Tomography}

\name{\em Mohammad Samin Nur Chowdhury$^{1}$, Diyu Yang$^{1}$, Shimin Tang$^{2}$, Singanallur V. Venkatakrishnan$^{3}$, \\ \em   Hassina Z. Bilheux$^{2}$, Gregery T. Buzzard$^{4}$, and Charles A. Bouman$^{1}$}
\vspace{-1in}
\address{$^{1}$School of Electrical and Computer Engineering, Purdue University, West Lafayette, IN 47907, USA\\ 
$^{2}$Neutron Scattering Division, Oak Ridge National Laboratory, Oak Ridge, TN 37830, USA\\ 
$^{3}$Electrical and Engineering Infrastructure Division, ORNL, Oak Ridge, TN 37830, USA\\ 
$^{4}$Department of Mathematics, Purdue University, West Lafayette, IN 47907, USA \thanks{This manuscript has been authored by UT-Battelle, LLC, under contract DE-AC05-00OR22725 with the US Department of Energy (DOE). The US government retains and the publisher, by accepting the article for publication, acknowledges that the US government retains a nonexclusive, paid-up, irrevocable, worldwide license to publish or reproduce the published form of this manuscript, or allow others to do so, for US government purposes. DOE will provide public access to these results of federally sponsored research in accordance with the DOE Public Access Plan (http://energy.gov/downloads/doe-public-access-plan).
}
}

\begin{document}
\setlength{\abovedisplayskip}{0pt} \setlength{\abovedisplayshortskip}{0pt}
\maketitle

\begin{abstract}
Hyperspectral neutron tomography is an effective method for analyzing crystalline material samples with complex compositions in a non-destructive manner. 
Since the counts in the hyperspectral neutron radiographs directly depend on the neutron cross-sections, materials may exhibit contrasting neutron responses across wavelengths. Therefore, it is possible to extract the unique signatures associated with each material and use them to separate the crystalline phases simultaneously.

We introduce an autonomous material decomposition (AMD) algorithm to automatically characterize and localize polycrystalline structures using Bragg edges with contrasting neutron responses from hyperspectral data. 
The algorithm estimates the linear attenuation coefficient spectra from the measured radiographs and then uses these spectra to perform polycrystalline material decomposition and reconstructs 3D material volumes to localize materials in the spatial domain. 
Our results demonstrate that the method can accurately estimate both the linear attenuation coefficient spectra and associated reconstructions on both simulated and experimental neutron data.
\end{abstract}


\section{Introduction}
\label{sec:introduction}
Hyperspectral neutron computed tomography (HSnT) is a powerful technique that has enabled 3D imaging of important material properties such as crystallographic phases \cite{woracek20143d} and isotopic compositions \cite{balke2021epithermal}.
HSnT is typically carried out at pulsed neutron source facilities by rotating a sample in the path of the beam, measuring the transmitted projection using a high-efficiency time-of-flight detector corresponding to each orientation, and processing the resulting data using a reconstruction algorithm. 
The combination of the pulsed source and time-of-flight detector enables resolving each projection into thousands of wavelength bins depending on the travel time of the measured neutrons \cite{josic2010energy}.

A typical workflow for HSnT involves processing the data corresponding to each wavelength bin using a reconstruction algorithm like filtered back projection (FBP), extracting the resulting spectra for each voxel in 3D \cite{woracek20143d}, and identifying homogeneous regions based on their spectral signature.    
For polycrystalline materials, these spectra correspond to the linear attenuation coefficient (or $\mu$-spectra) of a certain phase and are characterized by sharp dips in the attenuation values (called Bragg edges) at certain wavelengths. 
However, due to the low signal-to-noise ratio in each bin that is typical for HSnT data sets, the resulting reconstructions can be extremely noisy. 
Furthermore, it is inefficient to process each wavelength independently when the underlying sample is only composed of a few phases, each with its unique spectral signature. 

In order to address some of the challenges with existing approaches, dimensionality reduction methods have been applied in the context of HSnT. 
In \cite{balke2021epithermal, losko2022epithermal}, the authors have used a few known spectral signatures as a basis to decompose the measured HSnT data (in the epithermal range) and performed the reconstruction of this lower dimensional dataset, thereby significantly reducing the computation when compared to processing each of the wavelength bins independently. 
A recent work adopted a method of computing the $\mu$-spectra for polycrystalline materials utilizing known material locations in the hyperspectral reconstructions and using them for material decomposition \cite{ametova2021thermal}. 
However, none of these approaches are fully automated, as they use some form of supervision (known spectra or material locations).

In this paper, we present a novel algorithm for autonomous material decomposition (AMD) on hyperspectral neutron data without prior information. 
A key component of our algorithm is a two-stage subspace extraction procedure that transforms high-dimensional hyperspectral measurements into low-dimensional subspace projections.  
Stage one maps into an intermediate dimensional subspace that is designed to promote accurate, meaningful segmentation while rejecting most of the inherent noise and simultaneously drastically reducing the computation needed for reconstruction. 
Stage two uses this segmentation to produce the actual material decomposition.
A second key component of AMD is an unsupervised clustering procedure that automatically segments the material regions used for $\mu$-spectra estimation.
We test on both simulated and measured data and provide quantitative and qualitative performance evaluations.

\section{ Methodology}
\label{sec: methodology}
\begin{figure}[t!]
\centering
\centerline{\includegraphics[width=8 cm]{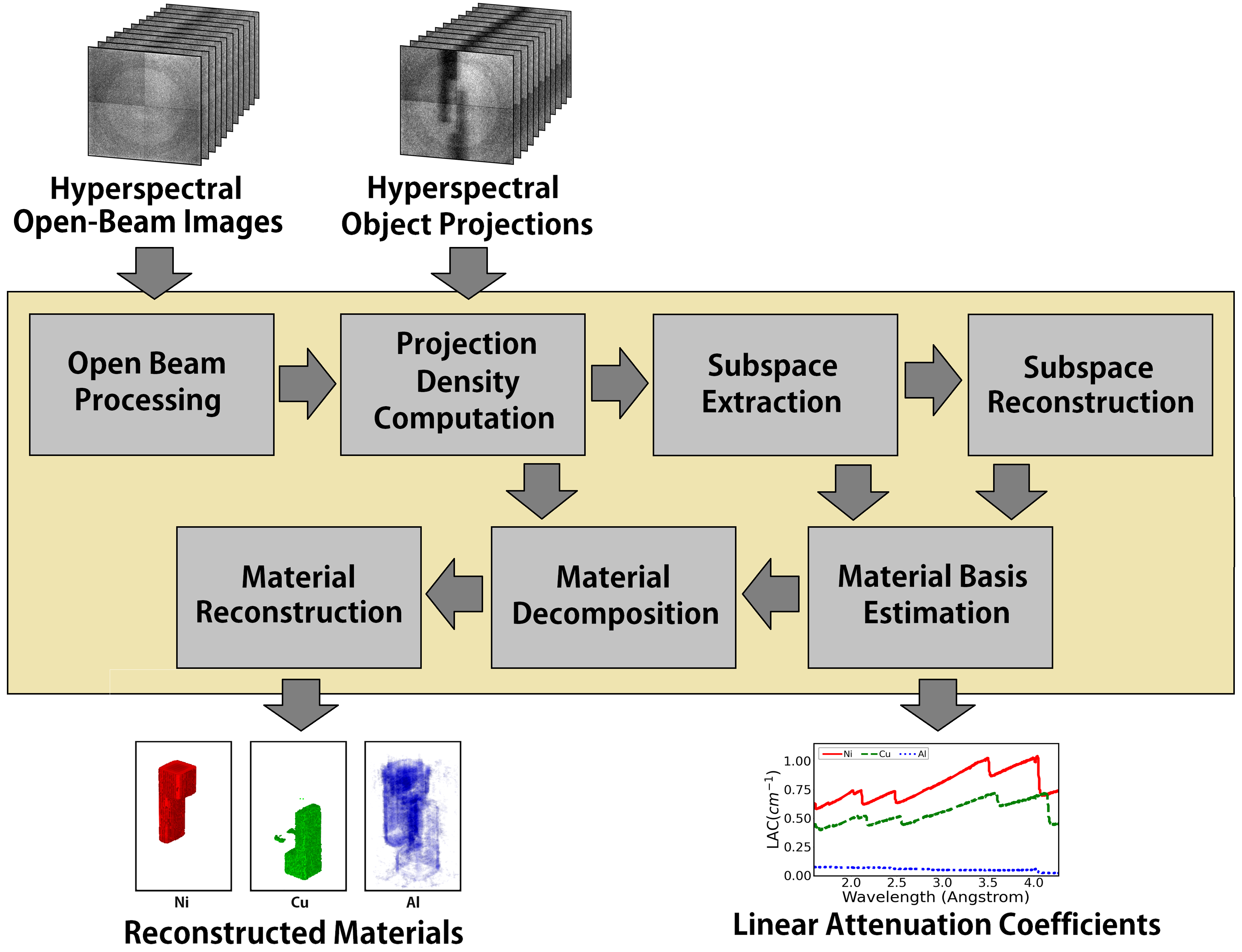}}
\vspace{-0.15in}
\caption{Illustration of the autonomous material decomposition (AMD) pipeline with sample inputs (hyperspectral neutron counts) and outputs ($\mu$-spectra, reconstructed materials).}
\label{fig:amd}
\end{figure}

Our AMD algorithm calculates $\mu$-spectra and material reconstructions from hyperspectral neutron radiographs. It uses low-noise open-beam images (also known as blank scans or flat fields) to normalize object projections. The algorithm decomposes these normalized hyperspectral projections using subspace basis vectors and reconstructs within this subspace, then performs clustering operations to spatially segment material regions. The individual $\mu$-spectra are calculated based on mean voxel values for each material region, after which the projection domain material decomposition and material reconstruction are performed. The entire procedure is illustrated in Figure \ref{fig:amd}.
\vspace{-0.09in}
\subsection{Open-Beam Processing}
\label{ssec:obproc}
To achieve the highest signal-to-noise ratio (SNR) possible for the open-beam images, we first average over multiple open-beam sets to reduce the noise. Then we filter the open-beam images at each wavelength using a 2D normalized Hamming kernel $h$ for further smoothing. This operation can be expressed as 2D convolution given by

\begin{equation}
	y_{r,c,k}^o = \sum_{i=-\infty}^{\infty} \sum_{j=-\infty}^{\infty} \bar{y}_{i,j,k}^o h_{r-i,c-j} \ ,
\end{equation}
where $y^o \in \mathbb{R}^{N_r \times N_c \times N_k}$ is the smooth open-beam array, $\bar{y}^o \in \mathbb{R}^{N_r \times N_c \times N_k}$ is the averaged open-beam array, $N_r$ is the number of rows in the detector, $N_c$ is the number of columns in the detector, and $N_k$ is the number of energies/wavelengths.
\vspace{-0.25in}
\subsection{Projection Density Computation}
\label{ssec:projdencomp}

To convert neutron counts to projection densities, we need to compute the negative log of the ratios of the object projections and the corresponding open-beam images. Ideally, the open-beam images $y^o$ would have the same neutron dosage rate as the object projections $y \in \mathbb{R}^{N_v \times N_r \times N_c \times N_k}$, where $N_v$ is the number of views. Then the ideal projection density array $p \in \mathbb{R}^{N_v \times N_r \times N_c \times N_k}$ can be expressed as
\begin{equation} \label{eq:p_ideal}
p_{v,r,c,k} = -\log\bigg( \frac {y_{v,r,c,k}} {y_{r,c,k}^o} \bigg) \ .
\end{equation}
However, in actual experiments, the dosage rate might vary for each view and energy, making the measured projection density $p^\text{meas} \in \mathbb{R}^{N_v \times N_r \times N_c \times N_k}$ deviate from the ideal. In order to correct for this, we subtract a view and wavelength-dependent offset $b \in \mathbb{R}^{N_v \times N_k}$ that makes the background approximately zero. This gives
\smallskip
\begin{equation}
p_{v,r,c,k} = p^\text{meas}_{v,r,c,k} - b_{v,k} \ .
\end{equation}
We estimate each $b_{v, k}$ by taking the spatial average of $p^\text{meas}_{v,k}$ in a region outside the sample, where $p_{v,k}$ is assumed to be 0. 
\vspace{-0.25in}
\subsection{Subspace Extraction}
\label{ssec:subextract}
To reduce the effect of noise and promote accurate material segmentation, we first use non-negative matrix factorization (NMF) \cite{pauca2006nmf} to reduce from full hyperspectral measurements to a lower dimensional subspace representation.  
More precisely, we seek to approximate $p = V D^t$, where $p \in \mathbb{R}^{N_p \times N_k}$ is the matrix obtained by using one hyperspectral projection density per row, $D\in \mathbb{R}^{N_k \times N_s}$ is a matrix whose columns are non-negative basis vectors for a subspace, and $V \in \mathbb{R}^{N_p \times N_s}$ gives the representation of each projection in subspace coordinates.  
Here $N_p = N_v \times N_r \times N_c$ is the number of projections, and $N_s$ is the dimension of the subspace.
Taking $N_m$ to be the number of materials, we choose $N_s$ so that $N_m<N_s<<N_k$. 
The resulting decomposition can be obtained by solving the optimization problem given by

\begin{equation}
	(V^{s}, D^{s}) = \argminX_{(V\geq 0, D\geq 0)} \{ \Vert p-V D^t \Vert_{F}^2 \} \ .
\end{equation}
Since $N_s<<N_k$, this reconstruction is significantly faster than full hyperspectral reconstruction.

\subsection{Subspace Reconstruction}
\label{ssec:subrecon}

Space domain analysis requires us to perform volumetric reconstructions. The subspace representation of $p$ allows us to perform high-quality reconstructions within a reasonable time using SVMBIR, a sophisticated Python package for super-voxel-based iterative CT reconstruction \cite{wang2016svmbir, svmbir-code}. 
The reconstructed subspace volume $x^s \in \mathbb{R}^{N_r \times N_c \times N_c \times N_s}$ is obtained by restricting to each subspace component, $x_j^s, j=1, \ldots, N_s$, one at a time and solving the associated optimization problem 
\medskip
\begin{equation} \label{eq:hatxjs}
    \hat x_j^s = \argminX_{x_j^s} \{ f(x_j^s) + h(x_j^s) \} \ ,
\end{equation}
where $f(x_j^s)$ is the forward model, and $h(x_j^s)$ is the prior model. SVMBIR is used to perform this optimization using Q-Generalized Gaussian Markov random field (QGGMRF) as the prior model \cite{pellizzari2017qggmrf} and a forward model of the form
\medskip
\begin{equation} \label{eq:forward}
    f(x_j^s) = \frac{1}{2 \sigma_v^2} \Vert V_j^s - Ax_j^s \Vert^2 \ ,
\end{equation}
where $A$ is the linear projection operator from a 3D volume to a rasterized sinogram, $V_j$ is the $j$th column of $V$, and $\sigma_v$ is the assumed standard deviation of the noise. 
\vspace{-0.09in}
\subsection{Material Decomposition and Reconstruction}
\label{ssec:matdec}

To convert from the subspace to a physically meaningful material basis, we first use GMCluster \cite{gmcluster}, a Gaussian mixture model-based clustering package, to segment the voxels in $x^s$ to identify regions corresponding to each material. Within each segmented region, we compute the vector mean to obtain a row of an $N_m \times N_s$ matrix $T$. With $T$, we have the relationship between subspace and material basis dictionaries given by
\smallskip
\begin{equation}
D^m = D^s T^t \ .
\end{equation}

We scale $D^m$ by a factor of $\frac{1}{\delta}$ to obtain the set of $\mu$-spectra, where $\delta$ is the voxel thickness/pixel pitch.

To perform material decomposition, we need to rerun NMF on the projection densities. However, this time, we calculate only $V^m$ using fixed $D^m$. The optimization problem is given by
\smallskip
\begin{equation}
	V^{m} = \argminX_{V\geq 0} \{ \Vert p-V(D^m)^t \Vert_{F}^2 \} \ .
\end{equation}
Finally, we estimate the desired material decomposition reconstruction $\hat x^m$ using
\medskip
\begin{equation} \label{eq:hatxjm}
    \hat x_j^m = \argminX_{x_j^m} \{ f(x_j^m) + h(x_j^m) \} \ ,
\end{equation}
where now $f$ is defined as in \eqref{eq:forward} with $V^m$ in place of $V^s$.

\vspace{-0.1in}
\section{Results}
\label{sec:results}
\vspace{-0.1in}
We implemented our AMD algorithm on both simulated and measured data. 
Each dataset contains neutron radiographs with 1200 wavelength bins ranging between 1.5 to 4.5 \AA.
We used $64\times64$ radiographs for simulated data and $512\times512$ for real data. 
The numbers of projection angles are 32 and 27 for simulated and measured data, respectively.

\begin{figure}[t!]
\centering
\centerline{\includegraphics[width=8.2cm]{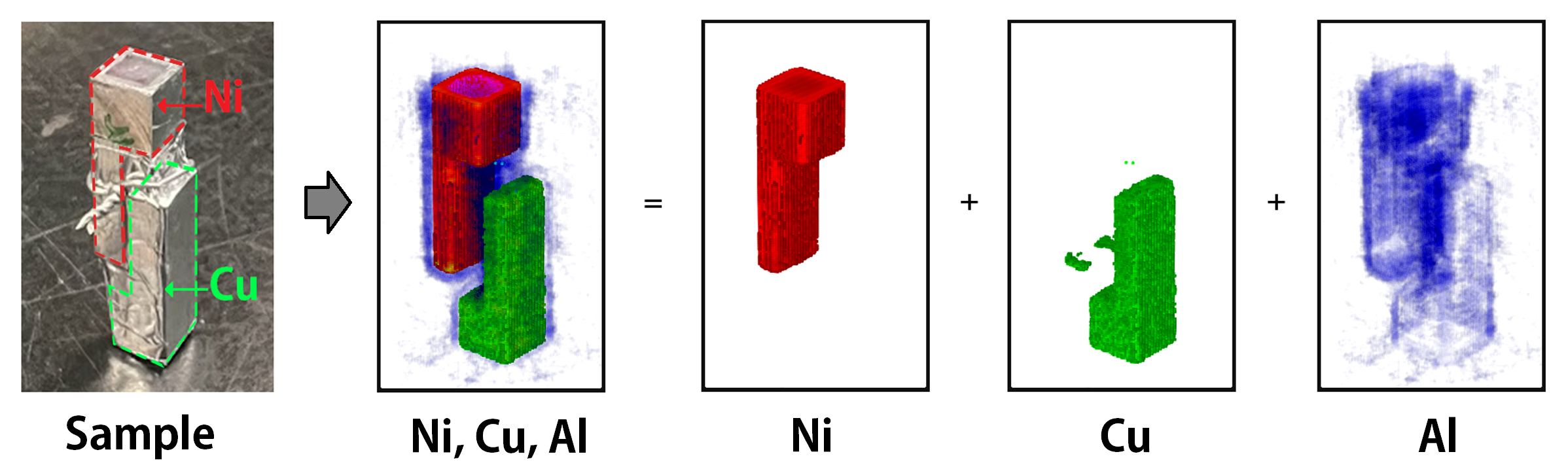}}
\vspace{-0.1in}
\caption{Reconstruction with measured data: Actual sample image (left) and 3D visualization of the reconstructed Ni, Cu, and Al estimated by AMD from measured data.}
\label{fig:recon_3D_REAL}
\end{figure}

The scanned physical phantom in Fig.~\ref{fig:recon_3D_REAL} is formed from powdered nickel (Ni) and powdered copper (Cu) in an aluminum (Al) structure. 
Data simulation was performed using a digital phantom that roughly approximated the physical phantom. 
The neutron measurements were then simulated using a Poisson noise model and attenuation calculations based on the object geometry and the theoretically known $\mu$-spectra computed using the Bragg-edge modeling (BEM) library \cite{lin2018bem}.
Heuristically, we found that 3 times the number of materials worked well as the subspace dimension for the cases we considered. 
So for all results, we used $N_m=3$ to represent the three materials (Ni, Cu, and Al) and $N_s=N_m \times 3=9$ for subspace extraction.
For the purpose of clustering, we considered the empty region surrounding the phantom as an additional material but discarded it later.

We compared our AMD results to a baseline in which each of the $N_k=1200$ wavelengths is reconstructed using the FBP method, after which the $\mu$-spectra are computed by averaging over a manually selected region of that material; we call this method reconstruction domain material decomposition (RDMD).
We note that RDMD is much more computationally expensive since it requires the computation of $N_k+N_m=1203$ reconstructions rather than the $N_s+N_m=12$ reconstructions required for our proposed AMD method.
Due to the high computational cost, we reconstructed only a single slice for each of the hyperspectral projections ($N_k$), while all slices for the material projections ($N_m$) in the implementation of the RDMD method.

\begin{figure}[t!]
\centering
\includegraphics[width=8cm]{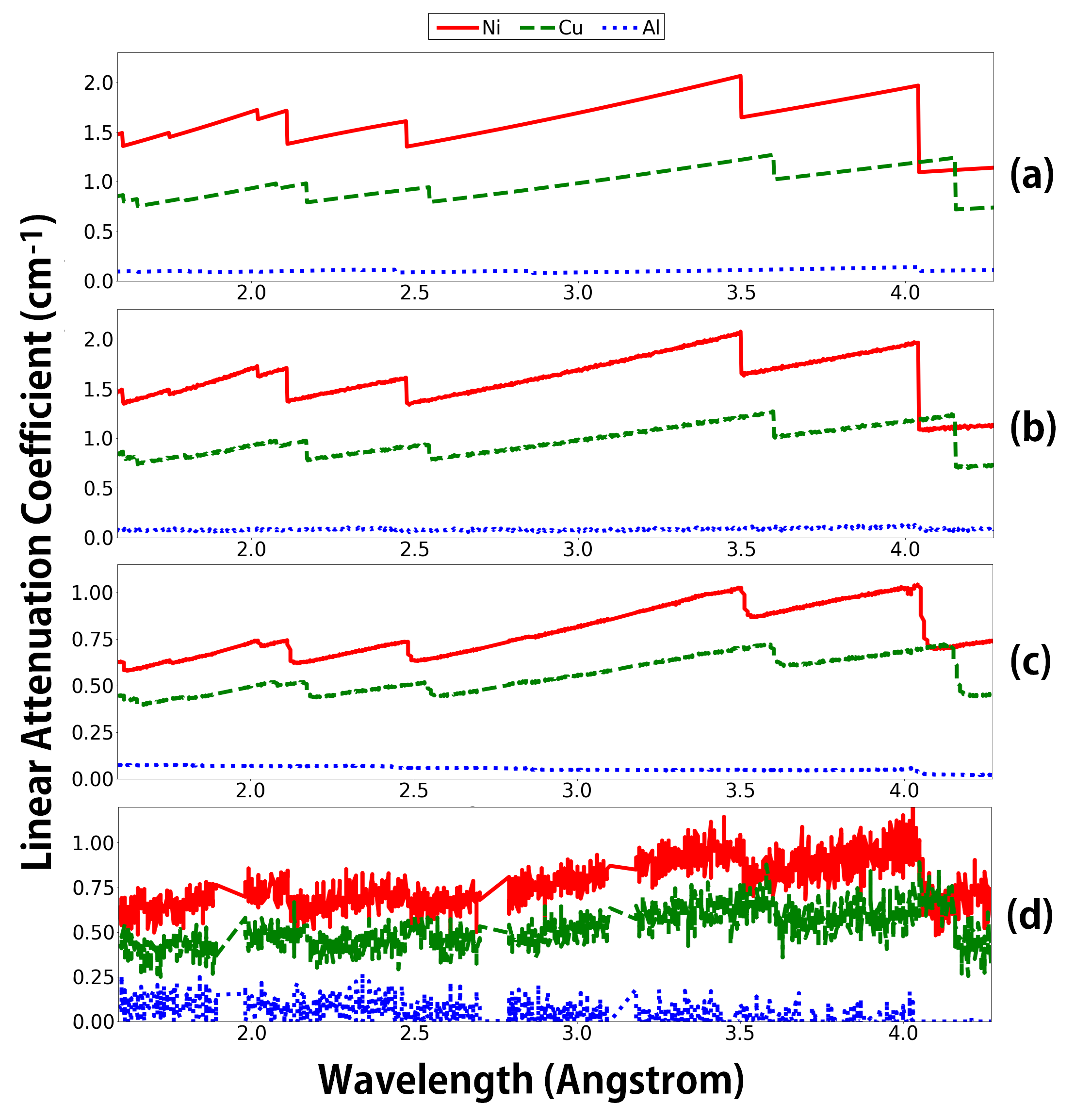}
\vspace{-0.1in}
\caption{$\mu$-spectra for Ni, Cu, and Al: (a) ground truth for simulation, (b) estimated by our algorithm (AMD) from simulated data, (c) estimated by AMD from measured data, (d) estimated by baseline (RDMD) from measured data.}
\label{fig:LAC_ALL}
\end{figure}

\begin{figure}[t!]
\begin{minipage}[a]{0.49\linewidth}
\centering
\includegraphics[width=4.4cm]{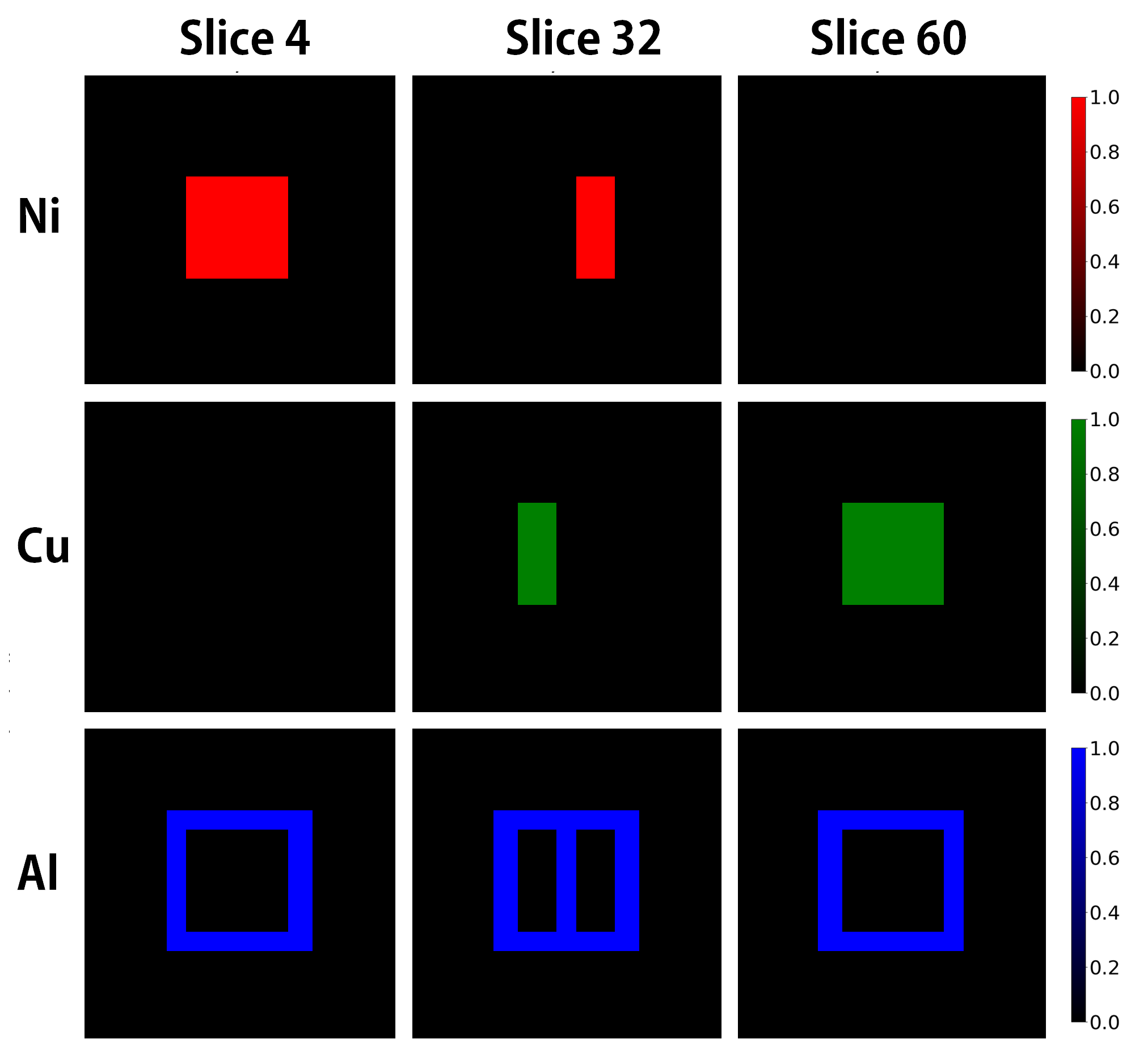}
\centerline{(a)}\medskip  
\end{minipage}
\hfill
\begin{minipage}[a]{0.49\linewidth}
\centering
\includegraphics[width=4.4cm]{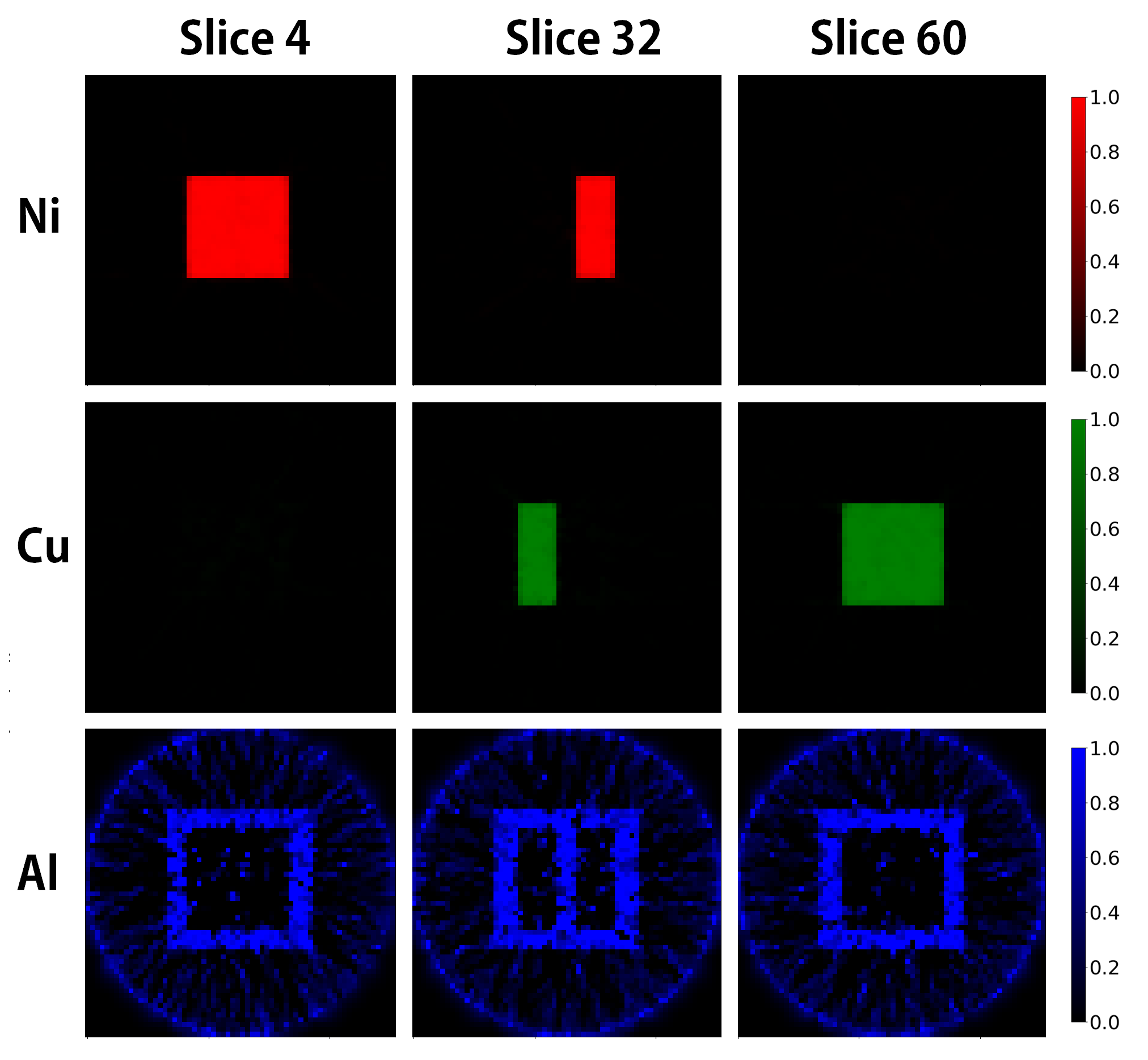}
\centerline{(b)}\medskip
\end{minipage}
\vspace{-0.1in}
\caption{Reconstruction with simulated data:  Selected slices for Ni, Cu, and Al in the space domain: (a)  ground truth for simulation, (b) estimated by AMD from simulated data.}
\label{fig:recon_SIM}
\end{figure}

\begin{figure}[t!]
\centering
\centerline{\includegraphics[width=8 cm]{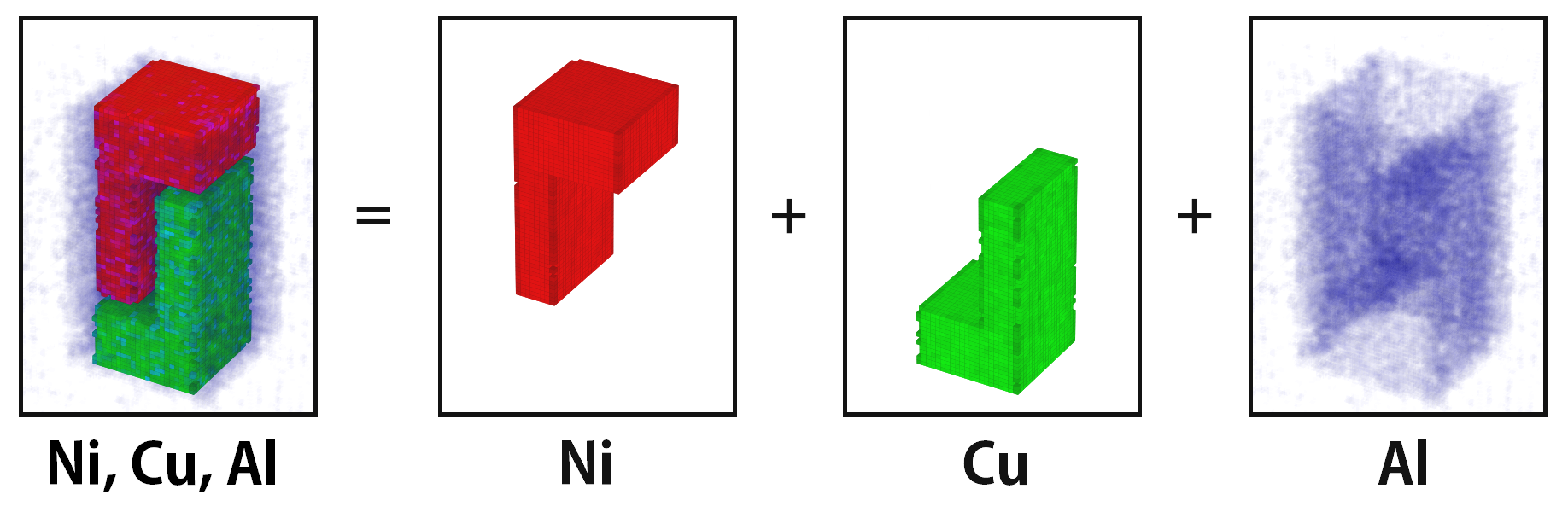}}
\vspace{-0.1in}
\caption{Reconstruction with simulated data:  3D visualization of the reconstructed Ni, Cu, and Al estimated by AMD from simulated data.}
\label{fig:recon_3D_SIM}
\end{figure}

Figure~\ref{fig:LAC_ALL} shows the results of $\mu$-spectra estimation for Ni, Cu, and Al in both the simulated and measured cases.
Fig.~\ref{fig:LAC_ALL}(a) shows the ground truth used in simulation;
Fig.~\ref{fig:LAC_ALL}(b) shows the spectra estimated from simulated data using AMD;
Fig.~\ref{fig:LAC_ALL}(c) shows the spectra estimated from measured data using AMD;
and Fig.~\ref{fig:LAC_ALL}(d) shows the spectra estimated from simulated data using the alternative RDMD method.

The AMD estimates of $\mu$-spectra accurately match the theoretical ground truth in the simulated case with an NRMSE of under 1\% for Ni and Cu; and match up to a scaling factor in the measured case, which we believe may be due to compaction of the powdered material; this scaling factor also appears in the RDMD results.
The AMD method extracts much cleaner and more accurate curves than the RDMD method.
In particular, the Bragg edge locations in the AMD estimates of $\mu$-spectra closely match the ground truth.

\begin{figure}[t!]
\begin{minipage}[a]{0.49\linewidth}
\centering
\includegraphics[width=4.4cm]{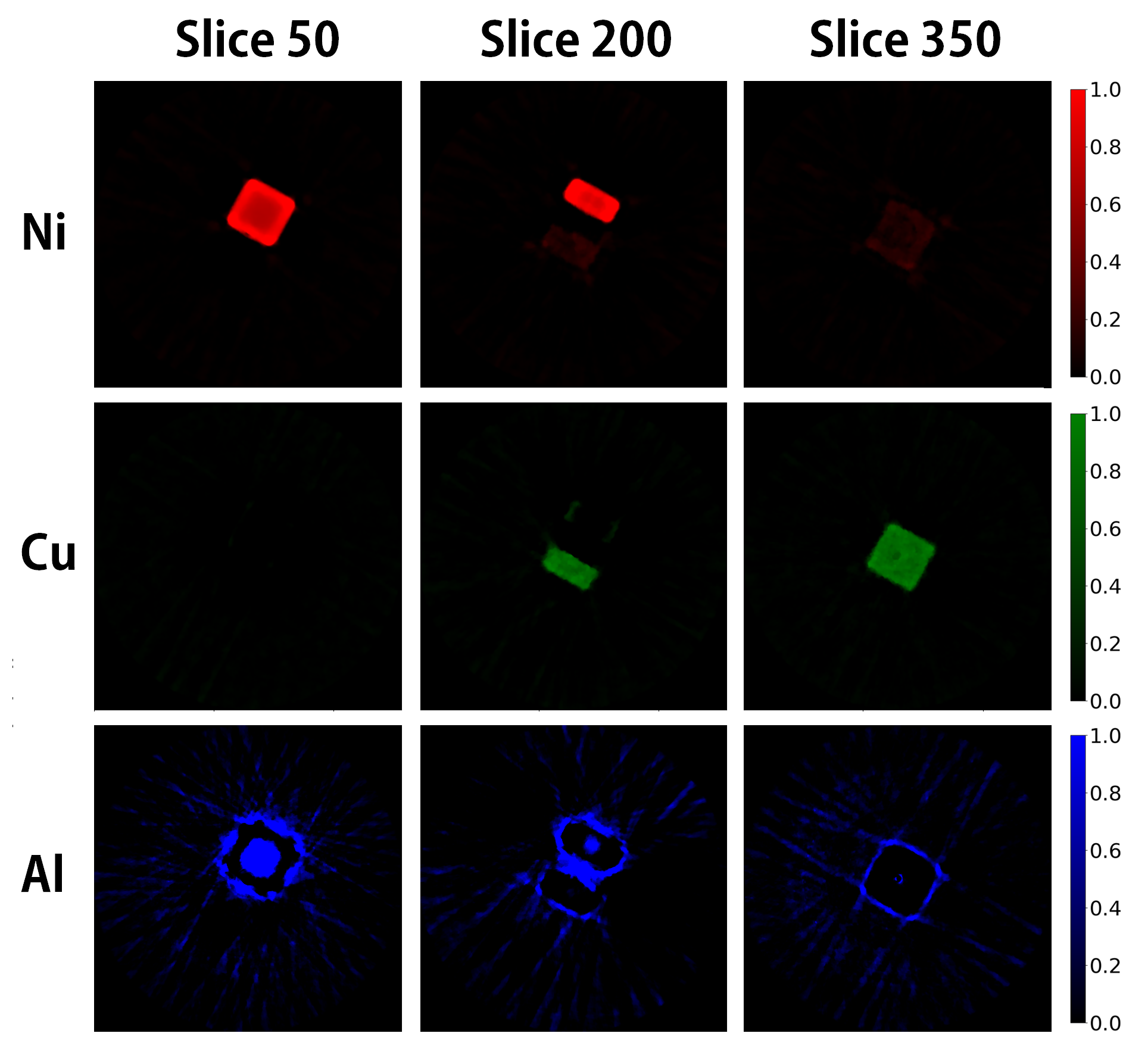}
\centerline{(a)}\medskip  
\end{minipage}
\hfill
\begin{minipage}[a]{0.49\linewidth}
\centering
\includegraphics[width=4.4cm]{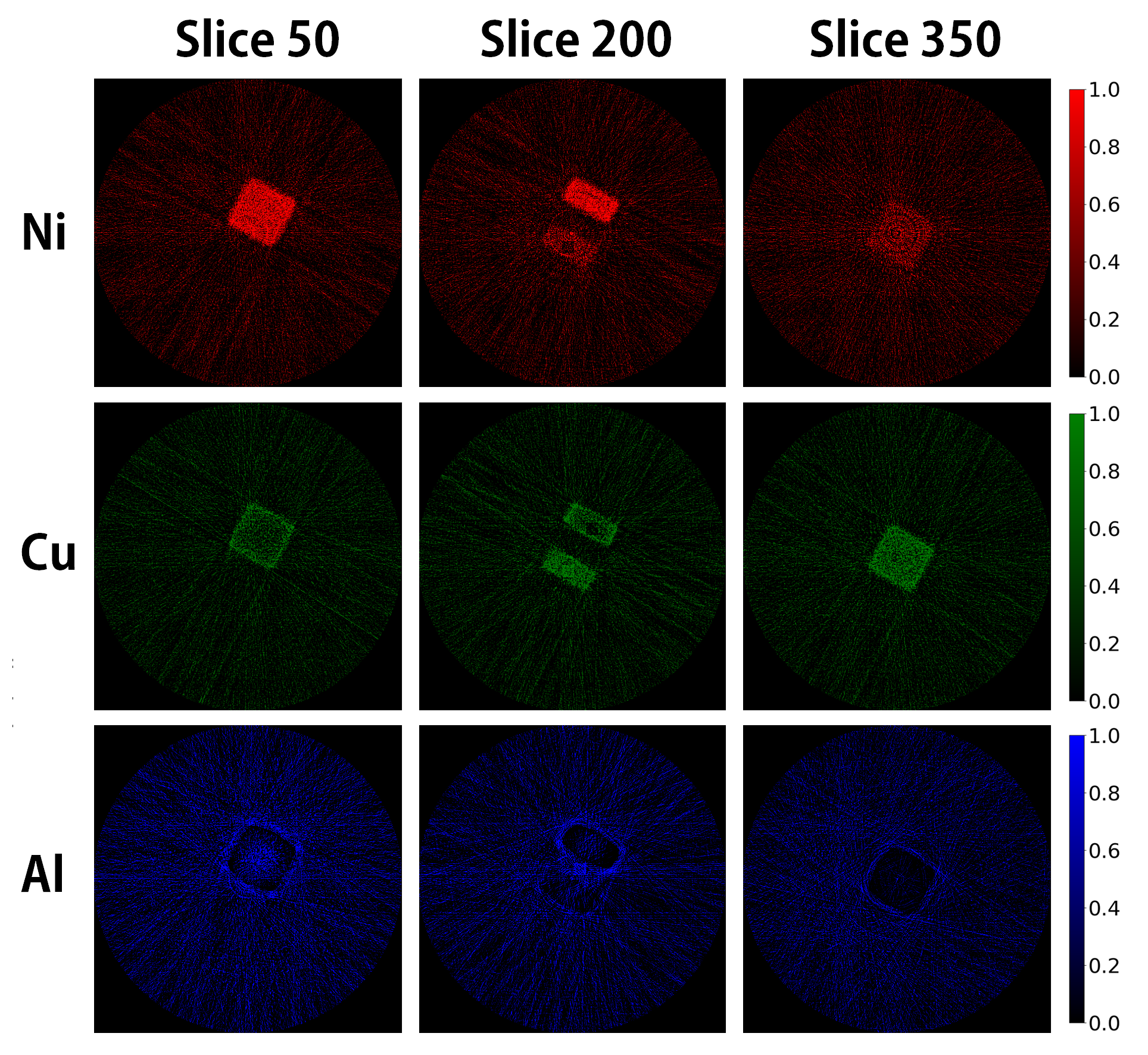}
\centerline{(b)}\medskip
\end{minipage}
\caption{Reconstruction with measured data:  Selected slices for Ni, Cu, and Al in the space domain (a) estimated by AMD from real data, (b) estimated by RDMD from real data.}
\label{fig:recon_REAL}
\end{figure}

Fig.~\ref{fig:recon_SIM}(a) shows some ground truth slices, and Fig.~\ref{fig:recon_SIM}(b) shows the associated AMD reconstructions from simulated data; Fig.~\ref{fig:recon_3D_SIM} shows 3D renderings of these reconstructions.
Notice that the AMD material decomposition of the phantom for simulated data is close to accurate.
Fig.~\ref{fig:recon_REAL}(a) and (b) show the reconstructions from measured data using  AMD and the alternative RDMD. The AMD results are substantially more accurate than the RDMD results, likely due to the more accurate estimates of the $\mu$-spectra. 
Again, Fig.~\ref{fig:recon_3D_REAL} shows a 3D rendering of the AMD reconstructions.
For both simulated and measured data, the artifacts in the Al reconstructions can be attributed to the weak presence of Al in neutron images making it difficult to separate Al from the background.

\vspace{-0.1in}
\section{Conclusion}
\label{sec: conclusion}
We introduce an autonomous material decomposition (AMD) method based on a 2-stage subspace extraction approach in the projection domain. 
The method identifies regions of homogeneous polycrystalline material compositions and extracts accurate linear attenuation coefficients spectra for each material resulting in accurate volumetric reconstructions.
By dramatically reducing the dimensionality of the hyperspectral data, AMD both reduces computation and improves the quality of reconstruction relative to baseline. 
While AMD is primarily devoted to Bragg edge imaging, we note that with some modifications, it may be useful for other hyperspectral modalities, such as neutron resonance imaging.

\section{Acknowledgment}
G. Buzzard was partially supported by NSF CCF-1763896, and C. Bouman was partially supported by the Showalter Trust. 
This research used resources at the Spallation Neutron Source, a DOE Office of Science User Facility operated by the Oak Ridge National Laboratory.
\newpage
\balance
\bibliographystyle{IEEEbib}
\bibliography{refs}
\end{document}